\documentstyle[11pt]{article}
\newcommand{\ket}[1]{\left | #1 \right \rangle}

\begin{document}

\begin{center}
{\large\bf QUANTUM EFFECTS IN ALGORITHMS} \bigskip \\ Richard
Jozsa\\ School of Mathematics and Statistics,\\ University of
Plymouth, Plymouth, Devon PL4 8AA, U.K.\\email:
rjozsa@plymouth.ac.uk\\[3mm]
 {\small\em For Proceedings of First NASA International Conference on
 Quantum Computation and Quantum Communication, Palm Springs, February 1998
 (appearing in a special issue of the journal  {\rm Chaos, Solitons
 and Fractals},  1998)}
\end{center}
\bigskip

\begin{abstract}
We discuss some seemingly paradoxical yet valid effects of quantum
physics in information processing. Firstly, we argue that the act
of ``doing nothing'' on part of an entangled quantum system is a
highly non-trivial operation and that it is the essential
ingredient underlying the computational speedup in the known
quantum algorithms. Secondly, we show that the watched pot effect
of quantum measurement theory gives the following novel
computational possibility: suppose that we have a quantum computer
with an on/off switch, programmed ready to solve a decision
problem. Then (in certain circumstances) the mere fact that the
computer {\em would} have given the answer {\em if} it were run, is
enough for us to learn the answer, even though the computer is in
fact {\em not} run.
\end{abstract}
\section{Introduction}
Many recent developments in quantum computation are motivated by
existing results in theoretical computer science, adapted and
rewritten in a quantum context. This includes much of the recent
work on quantum error correcting codes (see for example
\cite{er1,er2,er3}) and also the idea of using the Fourier
transform to determine periodicity, which underlies many of the
known quantum algorithms \cite{jozsa1}. There are relatively few
results (such as \cite{symm}) with no classical analogue, motivated
intrinsically from considerations of {\em physics}. This is a
curious situation considering that the entire subject of quantum
computation derives from differences between the classical and
quantum laws of physics. Apart from the computer science benefits
of providing more efficient computation, an important fundamental
aspect of the subject is the insight that it might provide for a
deeper understanding of the quantum laws and their origins.
Computer science and information theory provide an entirely new
conceptual framework for considering this question of physics. Thus
we will consider the question: what are the essential {\em
physical} effects that give rise to the known computational
speedups? And is it possible to use other differences between
quantum and classical physics for novel computational
possibilities?
\section{Quantum Information Processing and Entanglement}
It is often said that the power of quantum computation derives from
the superposition principle -- the ability to do different
computations in parallel in superposition, and combine the results
with cleverly arranged interferences. But this explanation is not
precise enough because classical waves also exhibit superposition
and any effect of superposition can be mimicked by a classical wave
system. However there is an essential difference between classical
and quantum superposition, which lies in the different way that the
two physical theories describe composite systems \cite{jgeom}.

Consider $n$ two-level systems. In the classical case we may for
example think of each system as comprising the two lowest energy
modes of vibration of a string with fixed endpoints together with
all superpositions. According to the laws of classical mechanics,
the total state space of the composite system is the {\em
Cartesian} product of the $n$ subsystem spaces. Thus no matter how
much the strings interact in their physical evolution, the total
state is always a product state of the $n$ separate systems. Hence
we can say that the information needed to describe the total state
grows {\em linearly} with $n$ (being $n$ times the information
needed to describe a single subsystem).

In contrast, according to the laws of quantum mechanics the total
space is the {\em tensor} product of the subsystem spaces and a
general state may be written as
\begin{equation} \label{psin} \ket{\psi_n} =
\sum_{i_1 , \ldots , i_n = 0}^{1} a_{i_1  \ldots  i_n} \ket{i_1}
\ldots \ket{i_n} \end{equation}
Thus generally we will have $O(2^n )$ superposition components
present and the information needed to describe the total state will
grow {\em exponentially} with $n$. The novel quantum effect here
-- the passage from Cartesian to tensor product -- is precisely the
phenomenon of entanglement i.e. the ability to superpose general
product states.

As stated above, quantum entanglement can be readily {\em mimicked}
by classical wave systems: instead of taking $n$ two-level systems,
we consider a single classical wave system with $2^n$ levels,
allowing general superpositions of all these levels, and merely
interpret these as entangled states via a chosen mathematical
isomorphism between $\otimes^n V_2$ and $V_{2^n}$ (where $V_k$ is a
$k$-dimensional vector space). However this mathematical
isomorphism is not a valid correspondence for considerations of
complexity (i.e. in which we assess the utilisation of physical
resources): if the $2^n$ classical levels are, say, equally spaced
energy modes, then to produce a general state in $V_{2^n}$ we will
need to invest an amount of energy exponential in $n$, whereas a
general state in $\otimes^n V_2 $ will require only a {\em linear}
amount of energy (as at most, each of the $n$ two-level systems
will need to be excited). The essential point here is that
entanglement allows one to construct exponentially large
superpositions with only linear physical resources and this cannot
be achieved with classical superposition.

In the sense described above the state $\ket{\psi_n}$ can encode an
exponentially large amount of information. This would be of little
consequence if we could not process the information in a suitably
efficient way. Fortunately the laws of quantum physics allow
precisely this possibility, which appears to be at the heart of the
computational speedup exhibited by the known quantum algorithms.
Suppose we apply a one-qubit gate $U$ to the first qubit of the
entangled state $\ket{\psi_n}$. This would count as just one step
of quantum information processing but to compute the result
classically (say by matrix multiplication) we would calculate the
new amplitudes by
\begin{equation} \label{uop} \tilde{a}_{j_1 \ldots  j_n} =
\sum_{i=0}^{1} U_{j_1 i} a_{ij_2 \ldots j_n} \end{equation}
where $U_{ji}$ is the unitary matrix for $U$. Now, this computation
involves exponentially many steps: the $2\times 2$ matrix
multiplication of $U$ needs to be performed successively $2^{n-1}$
times for all possible values of the indices $j_2 \ldots j_n$.
Although the action of $U$ on qubit 1 is a physically simple
operation, it is represented mathematically as a tensor product
$U\otimes I_2 \otimes \cdots \otimes I_2$ (where $I_2$ is the
identity matrix which represents ``doing nothing'' on qubits 2 to
$n$) and hence mathematically it becomes an exponentially large
unitary operation. Thus because of the tensor product rule we can
(somewhat enigmatically) state the principle:\\[1mm]{\bf (P1)}: The
physical act of doing nothing on part of an entangled composite
system is a highly nontrivial operation. It leads to an exponential
information processing benefit if used in conjunction with
performing an operation on another (small) part of the
system.\\[1mm] Indeed it is difficult to process the quantum
information by only a ``small amount''. Eq. (\ref{uop}) illustrates
that any small local operation (addressing a small part of the
system) will generally correspond to an exponentially large
processing operation from a classical point of view. Intuitively
this reflects the denseness of the exponential quantum information
stored within the linear resources.

One may object to {\bf (P1)}, claiming that surely the information
processing gain arises from the local operation that is actually
{\em performed} (e.g. $U$ above) rather than from the part that is
{\em not} performed (e.g. the $(n-1)$ identity operations above)!
To see that this is not the case consider our row of $n$ qubits and
suppose now that $U$ operates on the first $k$ qubits (so $U$ is a
$2^k \times 2^k$ matrix). Let us compare the number of steps
required to perform this transformation in the classical and
quantum contexts respectively. It is known that any $d \times d$
unitary matrix may be programmed on a quantum computer in $O(d^2 )$
steps \cite{d85,EJ} so the quantum implementation of $U$ will
require $O((2^k)^2 )$ steps. Classically, direct matrix
multiplication for a $d \times d$ matrix requires $O(d^2 )$ steps.
For $U$ we have $d=2^k$ and the multiplication must be performed
$2^{n-k}$ times. Thus the classical implementation will require $O(
(2^k )^2 2^{n-k})$ steps. Hence the ratio of quantum computing
effort to classical computing effort is $O(2^k / 2^n )$. This ratio
decreases if either $n$ is held fixed and $k$ is decreased, or $k$
is held fixed and $n$ is increased. In either case we are
increasing the proportion of ``doing nothing'' and this is giving
rise to an increased information processing benefit.

The Fourier transform is a fundamental ingredient
\cite{jozsa1,HO,CEM} in most of the known quantum algorithms which
exhibit a super-classical computational speedup. This includes the
algorithms of Deutsch \cite{DJ}, Simon \cite{SI}, Shor \cite{SH,EJ}
and Grover \cite{GR}. Using the mathematical formalism of the fast
Fourier transform (FFT) \cite{MASROC}, the unitary transformation
that is the Fourier transform can be implemented exponentially more
efficiently in a quantum context \cite{jozsa2} than in any known
classical context. For example, for the group  of integers modulo
$2^n$ the classical fast Fourier transform algorithm runs in time
$O(n2^n )$ whereas its quantum implementation runs in time $O(n^2
)$. An analysis of the implementation of the FFT algorithm in the
quantum context, given in detail in \cite{jozsa2}, shows that the
achieved exponential speedup may be entirely attributed to the
influence of the principle {\bf (P1)}. This appears to be an
essential feature of the speedup exhibited by all known quantum
algorithms.

The full (exponentially large) amount of information embodied in
the identity of a quantum state $\ket{\psi_n}$ is termed ``quantum
information''. The formalism of quantum mechanics places an
extraordinary limitation on the above entanglement-related benefits
of quantum information storage and processing: quantum measurement
theory implies severe restrictions on the accessibility of the
quantum information in the state. For example, according to
Holevo's theorem \cite{HOL} we can obtain at most $n$ bits of
information about the identity of an unknown state $\ket{\psi_n}$
of $n$ qubits by any physical means whatever. This bound is the
same as the information capacity of a classical system with the
same number of levels. Thus, curiously, natural physical evolution
in quantum physics corresponds to a super-fast processing of
(quantum) information at a rate that cannot be matched by any
classical means, but then, most of the processed information cannot
be read! It is a remarkable fact that these two effects do not
anull each other -- the small amounts of information that are
possible to obtain about the identity of the final processed state
do {\em not} coincide with the particular meagre kinds of
information processing that can be achieved by classical
computation on the input running for a similar length of time. This
disparity directly entails the computational speedup possibilities
of quantum computation.
\section{Counterfactual Quantum Computation}
We have argued above that the information processing benefits seen
in the known quantum algorithms all rest on some specific features
of quantum entanglement. However these features do not exhaust all
the ways in which quantum physics differs from classical physics.
In an effort to find new quantum algorithms we might ask whether
other non-classical features of quantum physics may be exploited
for novel computational possibilities (not necessarily just a
speedup of computation). Quantum measurement theory (c.f. the
inaccessibility of quantum information mentioned above) provides
further non-classical aspects of the quantum formalism and these
are also related to controversial interpretational issues. We will
now describe a novel computational possibility which we call
``counterfactual quantum computation'', based on properties of
quantum measurement.

A counterfactual effect may be defined as an observable physical
effect E whose outcome depends on an event A that might conceivably
have happened but in fact did not happen i.e. E is affected by the
mere existence of A as a valid possible alternative even though A
did not actually occur. Classical physics does not allow physically
observable counterfactual effects but quantum physics {\em does},
at least in the sense described below. Their surprising and
somewhat paradoxical occurrence in quantum mechanics has been
highlighted in Penrose \cite{penrose} (see especially \S\S
\hspace{0mm} 5.2, 5.3, 5.7, 5.8, 5.9, 5.18).

Suppose that we have a quantum computer which has been programmed
ready to solve a decision problem. The computer also has an on/off
switch, initially set in position off. We will show that in certain
circumstances, the mere fact that the computer {\em would} have
given the result of the computation {\em if} it were run, is
sufficient to cause a physically measureable effect from which we
can learn the result, even though {\em the computer is in fact not
run}! Our method is based on the so-called Elizur-Vaidman bomb
testing problem \cite{EV} and the essential idea may be clarified
by considering the operation of a simple Mach-Zender
interferometer, which we discuss first.

Consider the Mach-Zender interferometer as shown in the following
diagram.

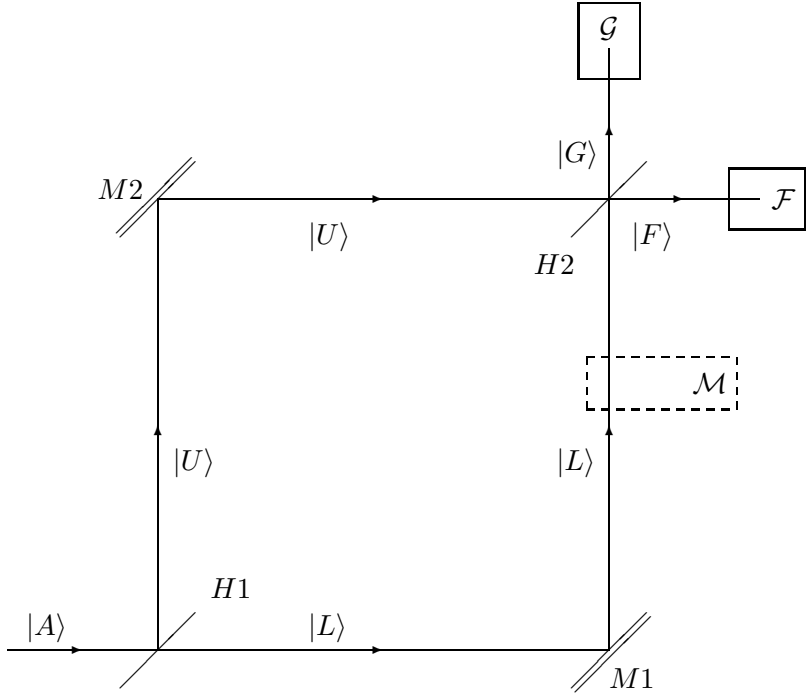
\begin{picture}(12,12)(0,0)
\put(1,2){\vector(1,0){1}} \put(2,2){\line(1,0){1}}
\put(3,2){\vector(1,0){3}}
\put(6,2){\line(1,0){3}}
\put(3,8){\vector(1,0){3}}
\put(6,8){\line(1,0){3}}
\put(3,2){\vector(0,1){3}}
\put(3,5){\line(0,1){3}}
\put(9,2){\vector(0,1){3}}
\put(9,5){\line(0,1){3}}
\put(9,8){\vector(0,1){1}} \put(9,9){\line(0,1){1}}
\put(9,8){\vector(1,0){1}} \put(10,8){\line(1,0){1}}
\put(1.25,2.2){$\ket{A}$}
\put(5,2.2){$\ket{L}$}
\put(3.2,4.4){$\ket{U}$}
\put(5,7.4){$\ket{U}$}
\put(8.3,4.4){$\ket{L}$}
\put(9.3,7.4){$\ket{F}$}
\put(8.3,8.5){$\ket{G}$}
\put(2.5,1.5){\line(1,1){1}}
\put(2.5,7.5){\line(1,1){1}}
\put(2.45,7.55){\line(1,1){1}}
\put(8.5,7.5){\line(1,1){1}}
\put(8.5,1.5){\line(1,1){1}}
\put(8.55,1.45){\line(1,1){1}}
\put(3.7,2.7){$H1$}  \put(2.2,8){$M2$}  \put(8,7){$H2$}  \put(9,1.5){$M1$}
\put(8.7,5.2){\dashbox{.1}(2,.7)[r]{${\cal M} \,\,$}}
\put(10.6,7.6){\framebox(1,.8)[r]{${\cal F} \,\,$}}
\put(8.6,9.6){\framebox(.8,1)[t]{\shortstack{\vspace{2pt}\\${\cal G}$}}}
\end{picture}
\\Here $H1$ and $H2$ are beam splitters and $M1$ and $M2$ are
rigid perfect mirrors. The action of each beamsplitter is taken to
be the following (written in terms of the states labelled at $H2$).
For horizontal photons \begin{equation} \label{u}
\ket{U} \rightarrow \frac{1}{\sqrt{2}} (\ket{F} + \ket{G}) \end{equation}
and for vertical photons \begin{equation} \label{l}
\ket{L} \rightarrow \frac{1}{\sqrt{2}} (\ket{F} - \ket{G}) \end{equation}
A photon enters at $\ket{A}$ and is separated into a superposition
$\frac{1}{\sqrt{2}}(\ket{L}+\ket{U})$ of upper and lower paths. In
the absence of the measuring instrument $\cal M$ the two beams
coherently interfere at $H2$ and according to eqs. (\ref{u}) and
(\ref{l}) the result is $\ket{F}$. Thus the photon is always
registered in detector $\cal F$ and never in detector $\cal G$.

Consider now a nondestructive measurement device $\cal M$ placed in
the lower arm, which registers whether or not the photon passed
along that arm. The initial state of $\cal M$ is $\ket{M_0}$ and if
a photon is registered it becomes an orthogonal state $\ket{M_1}$.
Following the photon we now have
\begin{equation} \ket{A} \rightarrow \frac{1}{\sqrt{2}}( \ket{U}+\ket{L})
\ket{M_0} \rightarrow \frac{1}{\sqrt{2}}( \ket{U}\ket{M_0} + \ket{L}
\ket{M_1}) \end{equation}
and the last state may be thought of as the ``collapsed'' mixture
of $\ket{U}\ket{M_0}$ or $\ket{L}\ket{M_1}$, each with probability
half. Thus the interference at $H2$ is spoilt and we always have a
50/50 probability of registering the photon in either $\cal F$ or
$\cal G$.

Suppose now that the photon is registered in $\cal G$ and the
measurement instrument is seen to be in state $\ket{M_0}$. (This
event occurs with probability $\frac{1}{4}$.) Thus the photon has
been registered absent in the lower arm and the measurement
instrument, having thus apparently done nothing, remains in state
$\ket{M_0}$. Yet the photon is seen at $\cal G$, which is forbidden
in the absence of $\cal M$! Although $\cal M$ apparently does
nothing, it cannot be removed, since then the photon can never
register in $\cal G$. This is our fundamental counterfactual
effect: we can say that the photon can be registered in $\cal G$
because {\em if} the photon would have gone along the lower path,
it {\em would} have been detected, even though it did not, in fact,
go along   the lower arm (since it was not seen by $\cal M$).

We can use this effect for computational advantage as follows.
Consider an idealised quantum computer which is an isolated
physical system containing an on/off switch, a set of program/data
registers denoted by the state $\ket{\rm comp}$ and an output
register. The on/off switch is a two-level system with basis states
$\ket{\rm on}$ and $\ket{\rm off}$ and the output register is a
two-level system with basis states $\ket{0}$ and $\ket{1}$. The
program/data registers are set up (``programmed'') to solve some
given decision problem together with its input (e.g. it might be
programmed to test for primality together with a given input
integer.) The output register, initially in state $\ket{0}$ will be
set by the computation to $\ket{0}$ or $\ket{1}$ according to the
answer of the decision problem. The length $T$ of the computation
is a known function of the input. The time evolution of the
computer for time $T$ is given by
\begin{eqnarray}
\ket{\rm on} \ket{\rm comp}\ket{0} \rightarrow \ket{\rm on}
\ket{\rm comp}\ket{r} \nonumber \\
\ket{\rm off} \ket{\rm comp}\ket{0} \rightarrow \ket{\rm off}
\ket{\rm comp}\ket{0} \nonumber \end{eqnarray}
Here $r=0$ or $1$ is the (initially unknown) result of the
computation and the computation will run only if the switch is set
to ``on''. The result is written into the output register and all
program/data registers are returned to their initial state.

Heuristically we will relate this scenario to the interferometer as
follows. $\cal M$ is the quantum computer with $\ket{M_0}$ and
$\ket{M_1}$ being the states $\ket{0}$ and $\ket{1}$ of the output
register. The photon is the on/off switch and the two paths are
delayed by a time $T$ for the photon to eventually arrive at $H2$.
Thus if $r=0$ the running of the computation makes no distinction
between the paths and the photon is always seen in $\cal F$. If
$r=1$ the computation (if it ran) would distinguish the two paths
and we will see the photon at $\cal G$ with probability
$\frac{1}{2}$. As before, with probability $\frac{1}{4}$ the photon
will register at $\cal G$ (so that we are sure that $r=1$) and the
output register will be seen to be in state $\ket{0}$. Thus the
computation has not run, yet we have learnt the result!

More formally in terms of states of the computer, we first set the
on/off switch to the superposition:
\begin{equation} \label{sw}
\left( \frac{\ket{\rm off}+\ket{\rm on}}{\sqrt{2}} \right)
\ket{\rm comp} \ket{0}
\end{equation}
and then allow time $T$ (the computation time) to elapse yielding the
state
\begin{equation}  \label{compstep}
\frac{1}{\sqrt{2}} \left( \ket{\rm off}\ket{\rm comp}\ket{0}+
         \ket{\rm on}\ket{\rm comp}\ket{r} \right)
\end{equation}
Next rotate the state of the switch by
\[  \ket{\rm off} \rightarrow \frac{1}{\sqrt{2}}(\ket{\rm off}+\ket{\rm on})
\hspace{8mm}
\ket{\rm on} \rightarrow \frac{1}{\sqrt{2}}(\ket{\rm off}-\ket{\rm on})
\]
This yields the state
\[
\frac{1}{\sqrt{2}}\left( \frac{(\ket{\rm off}+\ket{\rm on})}{\sqrt{2}}
\ket{\rm comp}\ket{0} +
\frac{(\ket{\rm off}-\ket{\rm on})}{\sqrt{2}}
\ket{\rm comp}\ket{r} \right)  \]
\begin{equation}  \label{final}
= \frac{1}{\sqrt{2}} \left( \ket{\rm off} \frac{(\ket{0}+\ket{r})}{\sqrt{2}}
+ \ket{\rm on} \frac{(\ket{0}-\ket{r})}{\sqrt{2}}\right) \ket{\rm comp}
\end{equation}
Here $r=0$ or 1 according to the (as yet unknown) result of the
computation. Next we measure the switch to see if it is on or off.
Note that if $r=0$ then we never see ``on'' and if $r=1$ we see
``on'' with probability $1/2$. Suppose that we see ``on''. Then we
know that the result of the computation must certainly be $r=1$. We
then examine the output register which will show $\ket{0}$ with
probability $1/2$. If this happens then {\em the computation has
not been run} (because if it had, then the output register must
show $\ket{1}$). Overall, if the result is actually $r=1$ then with
probability $1/4$ we learn the correct result (and know it is
correct) with no computation having taken place!

Note that if the actual solution of the decision problem is $r=0$
then we will never ascertain this from the above procedure because
if $r=0$ then the output register will always show 0 and the switch
will always be finally seen to be ``off''. But this outcome also
arises for $r=1$ with probability $\frac{1}{4}$ and we cannot {\em
a posteriori} distinguish the two possible causes. Correspondingly,
if the actual solution is $r=1$ then with probability $\frac{1}{4}$
will we fail to ascertain this.

The above description of the process represented by eqs. (\ref{sw})
to (\ref{final}) involves some delicate interpretational issues.
For example, a many-worlds adherent might object that initially the
switch was set in an equal superposition of being on and off, so
even in the subsequent case of ``no computation taking place'' the
computer actually did run in another ``parallel universe'' so we
cannot claim to get the result for free. One may, to some extent,
counter this objection as follows: suppose that when the result is
really $r=1$, the computer is also designed to explode at the end
of the computation, if it is run. Then using the above procedure,
in {\em my} world I learn that $r=1$ and the computer remains {\em
unexploded}, available to do another run. I do not really care if
it self-destructs in some ``other universe''!

The counterfactual quantum computation procedure above may be
considerably improved (using a method inspired by the improvements
to the Elizur-Vaidman problem given in \cite{kwiat}) to essentially
eliminate the deficencies noted above. As described below, we will
achieve the following:\\For any given $\epsilon > 0$
\begin{description}
\item[(i)] If the result is $r=0$, we will learn this with probability 1
but some computation will have taken place.
\item[(ii)] If the result is $r=1$, we will learn this with probability
$1-\epsilon$ with no computation having taken place.
\end{description} Thus for the many-worlds adherent, the universe in
which the computation takes place can be made to occur with {\em
arbitrarily small} amplitude $O(\sqrt{\epsilon})$ (in the case that
$r=1$), which considerably weakens his/her/its objection. Recall
that many basic results in information theory and computer science
are formulated in an asymptotic framework which allows an
arbitrarily small failure of some desired property. This occurs for
example in the distinction between the complexity classes P and BPP
\cite{PAP} (the latter allowing an arbitrarily small probability of
a false result) and Shannon's source coding theorem having not
perfect fidelity, but fidelity $1-\epsilon$ (for any $\epsilon
>0$) for the signals reconstructed from their coded compressed
versions. Thus if some undesirable result can be made to occur with
arbitrarily small (although non-zero) probability then FAPP it may
be ignored. \cite{fapp}

The improved counterfactual scheme exploits the so-called quantum
watched pot effect (or quantum Zeno effect) and it goes as follows.
We note first that the state $\ket{\rm comp}$ will never become
entangled with the other registers so we omit it, writing the
action of the computer as
\begin{equation}
\begin{array}{lll} \ket{\rm off}\ket{0}  & \rightarrow & \ket{\rm off}\ket{0}
\\
\ket{\rm on}\ket{0} & \rightarrow  & \ket{\rm on}\ket{r}
 \end{array} \end{equation}
Choose an angle $\alpha = \frac{\pi}{2N}$ for $N$ sufficiently
large (c.f. later). Then perform the following five operations:
\begin{description}
\item[(a)] Rotate the switch by angle $\alpha$.
\item[(b)] Allow the running time $T$ to elapse.
\item[(c)] Read the output register. If it shows 0 then continue.
If it shows 1 then discard the state and start again from the
beginning.\\{\em Remark.} (a) and (b) will result in the evolution
\begin{equation} \ket{\rm off}\ket{0} \rightarrow (\cos \alpha \ket{\rm off}
+\sin \alpha \ket{\rm on}) \ket{0} \rightarrow \cos \alpha \ket{\rm
off} \ket{0} + \sin \alpha \ket{\rm on}\ket{r}
\nonumber \end{equation} If $r=0$ then the output will {\em always}
show 0 and (c) will result in the state $(\cos \alpha \ket{\rm off}
+ \sin \alpha
\ket{\rm on} ) \ket{0}$ with probability 1. If $r=1$ then (c) will
result in the collapsed state $\ket{\rm off}\ket{0}$ obtained with
(high) probability $\cos^2 \frac{\pi}{2N}$. To complete the
procedure we:
\item[(d)] Repeat (a), (b) and (c) a further $N-1$ times.
\item[(e)] Finally measure the switch to see if it is on or off
(assuming that all stages have been kept in (c) and (d)).
\end{description}
We claim that in (e), if the switch is seen to be ``on'' then $r$
is certainly 0 (and some computation has been done), and if the
switch is seen to be ``off'', then $r$ is certainly 1 and no
computation has taken place. In the latter case the probability of
keeping all stages is $(\cos^2 \frac{\pi}{2N})^N$ which tends to 1
as $N\rightarrow \infty$. Thus by choosing $N$ to be sufficiently
large we can make the probability of success greater than
$1-\epsilon$ for any given $\epsilon$.

To see that our claim is correct, note that if $r=0$ then the
switch is just successively rotated from $\ket{\rm off}$ to
$\ket{\rm on}$ in $N$ stages and it never entangles with the output
register. If $r=1$ then the state is repeatedly collapsed to
$\ket{\rm off}\ket{0}$ so that no computation takes place in any
stage (because if it did, the output register would show the result
$r=1$). Indeed the waiting in (b) acts as a measurement of ``on''
versus ``off'' for the switch (if $r=1$) and in this case, we are
just freezing the switch in its $\ket{\rm off}$ state by frequent
repeated measurement. This is the quantum watched pot effect.

Note that according to {\bf (i)}, if $r=0$ then this result is not
learnt ``for free''. A natural question is whether or not there is
a counterfactual scheme which yields the information of {\em
either} result ($r=0$ or 1) with no computation having taken place.
The procedure described above may be readily modified to provide a
scheme with the following properties: with probability $1-\epsilon$
we learn the result and for either outcome, be it $r=0$ or $r=1$,
it is obtained for ``free'' with probability
$\frac{1-\epsilon}{2}$. We also learn whether or not the produced
result was obtained for ``free''. It remains an open question
whether or not each of the two results may be obtained for ``free''
with high probability $1-\epsilon$, or indeed, whether the sum of
these two probabilities can be made to exceed 1.

\end{document}